\documentstyle[12pt]{article}
\newcommand{\sbody}[2]{{\textstyle\frac{#1}{#2}}}
\newcommand{\stk}[1]{\stackrel{*}{\overline}}

\begin{document}
\begin{center}
\vfill
\large\bf{Antisymmetric Tensor Gauge Fields on $S_4$}
\end{center}
\vfill
\begin{center}
D.G.C. McKeon\\
Department of Applied Mathematics\\
University of Western Ontario\\
London\\
CANADA\\
N6A 5B7\vspace{.3cm}\\
\end{center}
\vfill
email: dgmckeo2@uwo.ca \hfill PACS: 11.30Ly\\
fax: 519-661-3523\\
tel: 519-661-2111, ext. 88789
\eject

\section{Abstract}

Antisymmetric tensor gauge fields $\phi_{ab}(\eta)$ are formulated on
the surface of a sphere $S_4\left(\eta^2 = a^2\right)$ embedded in five
dimensions. The free field model is equivalent to a scalar model on this
sphere. Interactions with gauge fields are discussed.  It is feasable to
formulate models for interactions with $U(1)$ gauge fields $A_a(\eta)$
that are akin to those of Freedman and Townsend in flat space. In
addition, it proves possible to have a novel interaction of $\phi_{ab}$
with $A_a$ and a spinor field $\Psi(\eta)$ on $S_4$ with both Abelian
and non-Abelian gauge invariance. In these models, $A_a$ plays the role
of a Stueckelberg field.

\section{Introduction}

The antisymmetric field $\phi_{\mu\nu}(x)$ has been of interest for some
time [1-4]. In flat space, the free field possesses the gauge invariance
$$\delta\phi_{\mu\nu} = \partial_\mu \theta_\nu -
\partial_\nu\theta_\mu\,;\eqno(1)$$
preserving this gauge invariance when $\phi_{\mu\nu}$ is coupled to a
vector gauge field is a problem that has been addressed by Freedman and
Townsend [5].

Vector gauge fields have also been considered on the surface of a sphere
$S_4\left(\eta^2 = \eta_a\eta_a = a^2\right)$ embedded in five
dimensional Euclidean space [6-12].  In this paper we consider how an
antisymmetric field $\phi_{ab}(\eta)$ can be defined on this surface and
how it can be coupled to a vector gauge field $A_a(\eta)$. The gauge
transformations that we will consider are of the form
$$\phi_{ab} \rightarrow \phi_{ab} + L_{ab}\theta\eqno(2)$$
$$A_a \rightarrow A_a + \eta_b L_{ba}\omega\eqno(3)$$
where $L_{ab} = \eta_a \partial_b - \eta_b \partial_a$. It will be shown
how models similar to those of ref. [5] can be formulated on $S_4$.

It will be also shown how a vector gauge field $A_a$ can serve as a
``Stueckelberg field'' for the tensor field $\phi_{ab}$. This will make
it possible to have self interactions for $\phi_{ab}$ and to have it
interact with a spinor field $\Psi$ without losing gauge invariance.
These interactions have a non-Abelian generalization. They do not appear
to have an analogue in flat space.

In Appendix A we review the results of ref. [5] and in Appexdix B some
features of gauge fields on $S_4$ are provided.

\section{Antisymmetric Tensor Field on $S_4$}

A suitable Lagrange density for a free antisymmetric tensor field
$\phi_{ab}$ on $S_4$ is
$$L = \sbody12 \Gamma_a\Gamma_a \;\;\;\;\left(\Gamma_a =
\epsilon_{abcde} L_{bc} \phi_{de}\right)\; .\eqno(4)$$
It is evident that this is invariant under the gauge transformation of
eq. (2) as well as the radial displacement
$$\phi_{ab} \rightarrow \phi_{ab} + \eta_a \sigma_b (\eta) -
\eta_b\sigma_a(\eta).\eqno(5)$$

It is possible to show that the Lagrange density in eq. (4) is
equivalent to
$$L = -\sbody25 \left[\phi_{ab}L_{pq}L_{pq}\phi_{ab} + 
\phi_{ab}L_{ab}L_{cd}\phi_{cd}\right. \eqno(6)$$
$$\left. -4 \phi_{pa}L_{ab}L_{bc}\phi_{cp}  + 8 \phi_{pa}L_{ab}\phi_{bp}\right].\nonumber$$

The Lagrange density of eq. (4) can be written in first order form
$$L = m\left(V_a\Gamma_a - \sbody12 m V_aV_a\right);\eqno(7)$$
this follows from substituting the equation of
motion for $V_a$
$$V_a = \frac{1}{m} \Gamma_a\eqno(8)$$
directly back into eq. (7).  The equation of motion for $\phi_{ab}$ is
$$\epsilon_{abcde} L_{cd} V_e = 0 \eqno(9)$$
which implies that $V_a$ is a pure gauge field
$$V_a = \eta_b L_{ba} \omega/m\; .\eqno(10)$$
This reduces the Lagrange density of eq. (7) to
$$L = -\sbody12 \left(\eta_\rho L_{pa}\omega\right)\left(\eta_q
L_{qa}\omega\right),\eqno(11)$$
which is equivalent to
$$L = \sbody12 \omega \left( \eta_p L_{ap}\right) \left(\eta_q
L_{aq}\right)\omega\eqno(12)$$
$$= \omega\left(\eta^2 L_{ab} L_{ab}\right)\omega .\eqno(13)$$
{\Big(}This is close to the Lagrange density for a massless scalar field
$\tilde{\omega}$ on $S_4$ given in ref. [6]
$$L = \eta^2\tilde{\omega}\left(L_{ab}L_{ab} - 4\right)\tilde{\omega}
.{\Big)}\eqno(14)$$
We have thus demonstrated how the Lagrange densities of eq. (4) and (7) have
properties analogous to those of eqs. (A.1) and (A.6).

We now consider a non-Abelian generalization of eq. (7),
$$L = -m\left(\sbody13 \epsilon_{abcde} \phi_{ab}^i F_{cde}^i(V) +
\frac{m}{2} V_a^i V_a^i\right)\eqno(15)$$
where now $i$ is an $SU(2)$ index and $F_{cde}^i(V)$ is defined by eq.
(B7).  This Lagrange density is invariant under the transformation
$$\phi_{ab}^i \rightarrow \phi_{ab}^i + L_{ab} \theta^i + \epsilon^{ijk}
\phi_{ab}^j \theta^k\eqno(16a)$$
$$V_a^i \rightarrow V_a^i + \epsilon^{ijk} V_a^j \theta^k\;
.\eqno(16b)$$
(Normally, for $F_{abc}^i(V)$ to transform as an isovector, one requires
the gauge transformation of $V_a^i$ to take the form of eq. (B8).
However, in (15) $F_{abc}^i(V)$ enters only through $\epsilon_{abcde}
F_{cde}^i(V)$ and
$$\epsilon_{abcde} L_{cd} \left(\eta_p L_{pe}\theta^i\right) =
0;\eqno(17)$$
it is thus sufficient to consider eq. (16b) to ensure that $L$ in eq.
(15) is an isoscalar.)

We now introduce a $U(1)$ vector gauge field $A_a$ and couple it to $\phi_{ab}$
using a Lagrange density analogous to eq. (A.14). The Lagrange density
$$L = m\left[\epsilon_{abcde} \left(A_a + V_a\right) L_{bc}\phi_{de} -
\sbody12 m V_a V_a\right]
-\frac{1}{12} F_{abc} (A) F_{abc}(A)\eqno(18)$$
(with $F_{abc}(A)$ given by eq. (B2)) possesses the invariances of eqs.
(2), (5), (B.4) and (B.5). The equation of motion for $\phi_{ab}$
$$\epsilon_{abcde} L_{cd} \left(V_e + A_e\right) = 0\eqno(19)$$
shows that $V_a + A_a$ is a pure gauge field so that
$$V_a = -A_a + \frac{1}{m} \eta_b L_{ba} \lambda .\eqno(20)$$
Substitution (20) into (18) gives rise to the $S_4$ analogue of the
Stueckelberg Lagrange density of eq. (A.17)
$$L = - \frac{1}{12} F_{abc} (A) F_{abc} (A) - \frac{m^2}2 \left(A_a -
\frac{1}{m} \eta_b L_{ba} \lambda \right).\eqno(21)$$
The gauge invariance of eq. (B4) is retained in eq. (21) provided
$$\lambda \rightarrow \lambda + m \omega\eqno(22)$$
though the invariance of eq. (B.5) is lost.

It is not immediately apparent how to generalize the Lagrange density of
eq. (18) so that one has a non-Abelian gauge invariance.

There is an essential difference between the gauge transformation for
the antisymmetric tensor field in flat space (eq. (11)) and on $S_4$
(eq. (2)). In the former case, the gauge function is itself a vector
while in the latter case it is a scalar. This suggests that the gauge
parameters for $A_a$ and $\phi_{ab}$ could be the same, and that the field $A_a$ could itself play
the role of the function $\sigma_a$ in eq. (5). Indeed, the quantity
$$\Lambda_a = \epsilon_{abcde} \left(\eta_b A_c - \sbody14 \eta^2
\phi_{bc}\right)\phi_{de}\eqno(23)$$
is invariant under the gauge transformation,
$$A_a \rightarrow A_a + \eta_b L_{ba} \theta \eqno(24a)$$
$$\phi_{ab} \rightarrow \phi_{ab} + L_{ab} \theta \eqno(24b)$$
as is
$$\Pi_{abc} = \epsilon_{abcde}\left(\eta_d A_a - \sbody12 \eta^2
\phi_{de}\right).\eqno(25)$$
As a result, the Lagrange density
$$L = -\frac{1}{12} F_{abc} (A) F_{abc} (A) + m \left(V_a \Gamma_a -
\sbody12 m V_aV_a\right)\eqno(26)$$
$$ + g_\Lambda \Lambda_a \Lambda_a + g_\Pi
\left(\Pi_{abc}\Pi_{abc}\right)^2 + g_{\Lambda\Gamma} \Lambda_a
\Gamma_a\nonumber$$
is invariant under the transformation of eq. (24). We also note that eq.
(26) is invariant under shifts of eq. (B.5), but that of eq. (5) is lost
for any term involving $\Lambda_a$ or $\Pi_{abc}$. Coupling to a spinor
field $\Psi$ is possible if we supplement the 
$U(1)$ version of the Lagrange density of eq.
(B.10) with interactions of the form
$$L = \lambda \Psi^\dagger \alpha \cdot \eta \left(\alpha_a \alpha_b
\alpha_c \Pi_{abc}\right)\Psi .\eqno(27)$$
The algebra of the matrices $\alpha_a$ implies that
$\epsilon_{abcde}\alpha_a\alpha_b\alpha_c$ is proportional to
$\gamma_{ab}$, and hence eq. (27) is equivalent to
$$L = \lambda \Psi^\dagger \alpha \cdot \eta \gamma_{ab} \left(\eta_a
A_b - \sbody12 \eta^2 \phi_{ab}\right)\Psi\; .\eqno(28)$$
One could of course use other combinations of $\Lambda_a$ and
$\Pi_{abc}$ to devise gauge invariant interactions; the ones considered
above are likely the only ones consistent with
renormalizability.

It is easy to generalize (26) and (27) to the non-Abelian case. For the
group $SU(2)$, eqs. (15) and (B.6) can be used in conjunction with the
isoscalar
$$\Lambda_a = \epsilon_{abcde}\left(\eta_b A_c^i \phi_{de}^i - \sbody14
\eta^2 \phi_{bc}^i \phi_{de}^i\right)\eqno(29a)$$
and the isovector
$$\Pi_{abc}^i = \epsilon_{abcde}\left(\eta_d A_e^i - \sbody12
\eta^2 \phi_{de}^i \right)\eqno(29b)$$
to construct the Lagrange density
$$L = - \frac{1}{12} F_{abc}^i(A) F_{abc}^i(A) - m\left(\sbody13
\epsilon_{abcde} \phi_{ab}^i F_{cde}^i (V) + \frac{m}{2} V_a^i
V_a^i\right)\nonumber$$
$$+g_\Lambda \Lambda_a\Lambda_a + g_\pi \left(\Pi_{abc}^i
\Pi_{abc}^i\right)^2\; .\eqno(30)$$
This is invariant under the gauge transformations of eqs. (16) and
(B.8). Furthermore, eq. (28) can be generalized to give
$$L = \lambda \Psi^\dagger \alpha \cdot \eta \gamma_{ab} \tau^i
\left(\eta_a A_b^i - \sbody12 \eta^2 \phi_{ab}^i \right)\Psi\;
\eqno(31)$$
while maintaining $SU(2)$ non-Abelian gauge invariance. We note that in
(26), (28), (30) and (31), $A_a$ and $\phi_{ab}$ can be decoupled by
making the shift 
$$\phi_{ab} \rightarrow \phi_{ab} +
\frac{1}{\eta^2}\left(\eta_a A_b - \eta_b A_a\right),\eqno(32)$$
leaving us with self interactions for the field $\phi_{ab}$.

\section{Discussion}

We have been able to demonstrate the first order formalism in flat space
of ref. [5] for coupling a $U(1)$ tensor gauge field $\phi_{ab}$ to a
vector gauge field $A_a$ has a direct analogue on a sphere $S_4$.
Furthermore, it has proved possible to couple $\phi_{ab}$ to $A_a$ on
$S_4$ in a manner that has no immediate flat space analogue. In the
latter model, symmetry of $\phi_{ab}$ under the radial shift of eq. (5)
is lost and there is but a single gauge parameter shared by $\phi_{ab}$
and $A_a$ (see eq. (24)). By shifting $\phi_{ab}$, $\phi_{ab}$ and $A_a$
can be decoupled.  A non-Abelian extension of eq. (18) has not (as yet)
been devised.

The problems of quantizing and renormalizing these models are currently
being addressed.

\section{Acknowledgements}

NSERC provided financial support.
R. and D. MacKenzie made a useful suggestion.

\section{Appendix A$\;\;\;$The Tensor Field}

In flat space, the tensor field $\phi_{\mu\nu} (x) = -\phi_{\mu\nu}(x)$
(diag $g_{\mu\nu} = (+, -, -, -)$) with the Lagrange desnity [1-4]
$$L = - \sbody18 G_\mu G^\mu \;\;\left( G^\mu =
\epsilon^{\mu\alpha\beta\gamma}\partial_\alpha
\phi_{\beta\gamma}\right)\eqno(A.1)$$
possesses the gauge invariance of eq. (1). The equation of motion that
follows from (A.1) is
$$G_{\mu ,\nu} - G_{\nu , \mu} = 0 ;\eqno(A.2)$$
this, combined with the identity
$$G^\mu_{\; ,\mu} = 0\eqno(A.3)$$
implies that we can write
$$G_\mu = \partial_\mu\sigma \eqno(A.4)$$
with the scalar $\sigma$ satisfying the massless Klein-Gordon equation
$$\partial^2 \sigma = 0 .\eqno(A.5)$$

The action of eq. (A.1) can be rewritten using a first order formalism in which $G_\mu$
is directly coupled to an auxiliary field $V_\mu$ [5]. If we have
$$L = \sbody12 m\left(G_\mu V^\mu + m V_\mu V^\mu\right) ,\eqno(A.6)$$
then the equation of motion for $\phi_{\alpha\beta}$ implies that
$$\epsilon^{\alpha\beta\gamma\delta} \partial_\gamma V_\delta =
0\eqno(A.7)$$
so that
$$V_\mu = \partial_\mu \lambda /m^2\eqno(A.8)$$
reducing (A.6) to
$$L = +\sbody12 \left(\partial_\mu \lambda\right)^2; \eqno(A.9)$$
again demonstrating that the model of eq. (A.1) is just that of a free
spin zero field.  (A non-Abelian generaliztion of eq. (A.6) leads [5] to
a non-linear sigma model in place of eq. (A.9).)

If we use the equation of motion of $V_\mu$
$$V_\mu = \frac{-1}{2m} G_\mu\eqno(A.10)$$
to eliminate $V_\mu$ in eq. (A.6) then we recover eq. (A.1).

We can also supplement eq. (A.6) with a term proportional to
$\phi_{\mu\nu}\phi^{\mu\nu}$ so that
$$L = \frac{m}{2} \left(G_\mu V^\mu + m V_\mu V^\mu \right) - \kappa
\phi_{\mu\nu} \phi^{\mu\nu}\; .\eqno(A.11)$$
(This breaks the gauge invariance of eq. (1).) In place of eq. (A.7) we
now find
$$\phi_{\mu\nu} \sim \epsilon_{\mu\nu\alpha\beta} \partial^\alpha
V^\beta\eqno(A.12)$$
and with an appropriate choice of $\kappa$, eliminating $\phi_{\mu\nu}$
in eq. (A.11) leaves us with the Lagrange denisty for a massive vector
(Proca) field
$$L = -\sbody14 F_{\mu\nu} F^{\mu\nu} + \sbody12 m^2
V_\mu V^\mu\;.\eqno(A.13)$$

Coupling $\phi_{\mu\nu}$ to a $U(1)$ gauge field $A_\mu$ by
complexifying $\phi_{\mu\nu}$ and replacing
$\partial_\mu\phi_{\alpha\beta}$ by $\left(\partial_\mu -
ieA_\mu\right)\phi_{\alpha\beta}$ in eq. (A.1) breaks the symmetry of
eq. (1). Freedman and Townsend circumvent this by directly coupling
$G_\mu$ to $A_\mu$, using the Lagrangian density
$$L = \sbody12 m\left[G_\mu \left( V^\mu + A^\mu\right) + m V_\mu
V^\mu\right]\eqno(A.14)$$
$$ - \sbody14 F_{\mu\nu} (A) F^{\mu\nu} (A)\nonumber$$
where
$F_{\mu\nu} (A) = \partial_\mu A_\nu - \partial_\nu A_\mu$. (A 
non-Abelian generalization of eq. (A.14) also appears in ref. [5].) This
Lagrangian is invariant under the transformations of eq. (1) as well as
$$A_\mu \rightarrow A_\mu + \partial_\mu \omega\; .\eqno(A.15)$$
The equation of motion for $\phi_{\mu\nu}$ gives, in place of eq. (A.8),
$$V_\mu = -A_\mu + \frac{1}{m} \partial_\mu\lambda\eqno(A.16)$$
so that elimination of $V_\mu$ in eq. (A.14) results in the Stueckelberg
Lagrange density for a massive, gauge invariant vector field
$$ L = -\sbody14 F_{\mu\nu} (A) F^{\mu\nu} (A) + \frac{m^2}{2}
\left(A_\mu - \frac{1}{m} \partial_\mu\lambda\right)^2 .\eqno(A.17)$$
Invariance under eq. (A.15) is retained, provided
$$\lambda \rightarrow \lambda + m\omega.\nonumber$$

\section{Appendix B $\;\;\;\;$ Gauge Fields on $S_4$}

In refs. [7-12], vector gauge fields are formulated on $S_4$. In the $U(1)$
case, the Lagrange density for the vector $A_a$ is
$$L = - \frac{1}{12} F_{abc} (A) F_{abc} (A) \eqno(B.1a)$$
or equivalently,
$$L = \sbody12 A_a P_{ab} A_b\eqno(B.1b)$$
where
$$F_{abc}(A) = L_{ab} A_c + L_{bc}A_a + L_{ca}A_b\eqno(B.2)$$
and
$$P_{ab} = \sbody12 L_{pq} L_{pq}\delta_{ab} + L_{ap} L_{pb} - 3L_{ab}.\eqno(B.3)$$
The action of eq. (B.1a) is invariant under the gauge transformation
$$A_a \rightarrow A_a + \eta_b L_{ba} \omega \eqno(B.4)$$
as well as the radial shift
$$A_a \rightarrow A_a + \eta_a \lambda\; .\eqno(B.5)$$
A non-Abelian generalization of (B.1a) is [9, 10, 12]
$$L = -\frac{1}{12} F_{abc}^i F_{abc}^i \eqno(B.6)$$
where now for the $SU(2)$ case
$$F_{abc}^i = \left(L_{ab} A_c^i + \epsilon^{ijk} \eta_a
A_b^j A_c^k\right) + (\rm{cyc.\;\;perm.\;\;of\;\;}abc).\eqno(B.7)$$
Eq. (B.6) is invariant under the gauge transformation
$$A_a^i \rightarrow A_a^i + \eta_b L_{ba} \theta^i + \epsilon^{ijk} A_a^j
\theta^k\eqno(B.8)$$
as well as under the radial shift
$$A_a^i \rightarrow A_a^i + \eta_a \lambda^i.\eqno(B.9)$$

The gauge field $A_a^i$ can also be coupled to a spinor field
$\Psi$ [9, 12], using the Lagrange density
$$L = \Psi^\dagger \alpha_a \eta_a\left[- \gamma_{pq} L_{pq}
+ 2 + \gamma_{pq}\left(\eta_p A_q^i - \eta_q A_p^i\right)\tau^i\right]\Psi\eqno(B.10)$$
where
$$\left\lbrace \alpha_a , \alpha_b \right\rbrace = 2\delta_{ab}\eqno(B.11a)$$
$$\gamma_{ab} = \sbody14 \left[ \alpha_a , \alpha_b\right],\eqno(B.11b)$$
and $2\tau^i$ is an $SU(2)$ Pauli matrix. Gauge invariance is retained in eq. (B.10).

\end{document}